\documentstyle[aps,prl,epsfig,floats]{revtex}

\def\gesim{\lower0.5ex\hbox{$\:\buildrel >\over\sim\:$}}
\def\lesim{\lower0.5ex\hbox{$\:\buildrel <\over\sim\:$}}
\def\gev{\hbox{GeV}}
\def\tev{\hbox{TeV}}
\def\qwe{f}

\def\slash{\!\!\!\!/}
\def\g2{g\!-\!2}
\def\gmu2{g_\mu\!\!-\!2}
\def\ocal{{\cal O}}
\def\lcal{{\cal L}}
\def\ie{{\it i.e.}}
\begin{document}

\preprint{MCTP-01-12 \\ UCRHEP-T298}
\twocolumn[\hsize\textwidth\columnwidth\hsize\csname @twocolumnfalse\endcsname

\title{Model-Independent Analysis of $g_\mu-2$}

\author{Martin B. Einhorn$^a$ and J. Wudka$^b$}
\address{$^a$ Randall Laboratory of Physics, University of Michigan,
Ann Arbor, Michigan, 48019-1120\\
$^b$ Department of Physics, University of California, Riverside CA 92521-
0413.}
\date{March 2, 2001}
\maketitle

\begin{abstract}
Abstract:  Assuming that the scale of new physics exceeds the weak scale, we
have considered all possible origins of deviations from the Standard Model in
$\gmu2$.  If the underlying theory can be treated perturbatively, then to
account for an effect as large as has been recently reported\cite{BNL}, the
only possibilities would be models that contain particles that yield direct
contributions, of which SUSY models are prototypical, or models having
unexpectedly light leptoquarks or having a rather exotic heavy vector boson
that couples predominantly to muons. Other possibilities either are already
excluded by existing experimental results or would have to be unnaturally large.
If, on the contrary, the underlying new physics involves strong interactions,
as in technicolor models, then the discrepancy could be accounted for by a
variety of effective operators.
\end{abstract}

\draft

%----------------------------------------------------------------------

\vskip1pc]

\paragraph{Introduction}

The announcement of a new, more precise measurement of the anomalous magnetic
moment of the muon\cite{BNL} that apparently disagrees with the Standard Model
(SM) prediction has stirred a great deal of excitement.  This hint of new
physics, coming on the heels of suggestions from LEP of a light Higgs
boson\cite{LEP}, supports a picture in which there are particles previously
undetected already at the scale of 100~\gev~ or so.  In this note, we wish to
evaluate the theoretical implications of the new result in a rather model-
independent manner, based on our earlier analysis\cite{aew}, hereinafter
referred to as AEW.  The method that allows for such an analysis is the
effective field theory approach in which one identifies the leading-order
higher-dimensional operators that might occur if departures from the SM are
due to physics on some high mass scale $\Lambda.$  We assumed that the new
physics respects the SM gauge symmetries. The operators therefore are simply
monomials involving fields of the SM.  By SM, we made the conventional
assumption that there was but a single Higgs doublet.  There is nothing in
existing data that requires this, but both the conventional SM ``prediction"
as well as our analysis would have to be modified if there were more than one
Higgs doublet.

There are essentially two different types of new dynamics to be associated
with the physics beyond the SM, which we referred to as ``decoupling" or
``non-decoupling," depending on whether the underlying dynamics involves weak
coupling or strong interactions.  In the former case, dimensional analysis may
be used to analyze new effects, while in the latter case, one must resort to
an expansion of transition amplitudes in powers of momenta, commonly referred
to as a derivative expansion or chiral perturbation theory.

In AEW, we assumed as a figure of merit that the BNL $\gmu2$ experiment would
achieve its intended accuracy of $\Delta a_\mu\simeq 4\times 10^{-10}.$  The
recent interim result\cite{BNL} is statistics limited, with an error of
$\Delta a_\mu=15\times 10^{-10}$.\footnote{After the experimenters analyze
their year 2000 data, their accuracy is expected to improve by about a factor
of two.} Nevertheless,  the central value observed differs from the SM
prediction by $\delta a_\mu= 42\times 10^{-10},$  about a factor of three
larger than the SM weak effect of $\delta a_{weak}\simeq 15\times 10^{-10}$.
Such a relatively large effect is at first sight rather surprising, given the
lack of evidence of deviations from the SM in the high-precision LEP
experiments\cite{LEPprecise} or elsewhere. Indeed, such a relatively large
effect will be seen to simplify the model-independent interpretation about the
origin of new physics.

\bigskip
\paragraph{Decoupling Case}

If physics is sufficiently weakly interacting so that perturbation theory may
be applied, the dynamical dimensions of fields are approximately equal to
their naive (or engineering) dimensions.  The SM itself is of this type (at
energies $E\gg\Lambda_{QCD}$) and moreover is ``natural" in the sense that all
gauge-invariant operators of dimension four or less are presumed to occur.
Assuming conservation of baryon and lepton number, the first corrections to
the SM are operators of dimension six.  These were catalogued long-
ago\cite{Buchmuller:1986jz}, where it was shown that there were 81 independent
such operators.\footnote{The catalogue is also given in Appendix~A of
Ref.~\cite{patterns}.} Although their number may seem daunting, for any given
process or measurement, usually only a small subset of these are relevant.  We
write them in the generic form $ \alpha \ocal/\Lambda^2$, where $\ocal$ is
some gauge-invariant monomial of dimension six constructed from SM fields and
their covariant derivatives, and $\Lambda$ is referred to as the scale of new
physics. To be slightly more precise, the scale $\Lambda$ is to be compared
with the weak scale $v\approx 250~\gev$ and not to be associated directly with
some virtual particle's mass.\footnote{One usually speaks of a single scale
$\Lambda$ but of course there may actually be several sources of new physics
characterized by very different scales.} The dimensionless coupling constant
$\alpha$ depends on some renormalization scale, which we tacitly take to be
$\Lambda$.  In principle, the coupling should be run down to some small
momentum scale, but, since we will be only making order of magnitude estimates
in this paper, we will ignore this running.

It is important to understand the natural size of the coefficient $\alpha$ of
each of these dimension-six operators.  That is facilitated by classifying
them into those that may arise from tree diagrams in any underlying theory and
those that can only arise from loop diagrams. Those arising from loops are
expected to be suppressed relative to tree diagrams by at least one additional
power of {$1/16\pi^2$} times some coupling constant(s).  Such a classification
was carried out in Ref.~\cite{patterns}, to which we shall refer as needed.

For the case at hand, we may conveniently classify the operators into two
types, those that contribute directly to a change in $\g2$ at the tree level
and those that contribute only indirectly through their presence in loop-
graphs.\footnote{The reader unfamiliar with effective field theory may be
troubled by the inclusion of ``nonrenormalizable" operators in loop
corrections, but such theories, containing as they do an {\em infinite} number
of higher dimensional operators, are in fact sensible and interpretable as
quantum field theories. For further discussion, see for example
Weinberg\cite{weinberg}.} Nearly all dimension-six operators contribute only
indirectly to $\g2$. There are just two operators\footnote{See AEW for
details.  In fact, one might say there is only one such operator, since one
linear combination of these two will decouple from the photon.} that
contribute directly to changes in $\g2$, viz.,
\begin{eqnarray}
\ocal_{ \mu W } &\equiv& ( \bar l_\mu \sigma^{\alpha\beta} \tau^I \mu_R )
\phi W_{\alpha\beta}^I + h.c.\cr
%\quad \hbox{and}
\quad \ocal_{\mu B }&\equiv&( \bar l_\mu \sigma^{\alpha\beta} \mu_R )
\phi B_{\alpha\beta} + h.c.,
\label{directo}
\end{eqnarray}
where $l_\mu$ denotes the left-handed doublet of the SM containing the $\mu^-
_L$ and $\nu_\mu$ fields.  Referring to Ref.~\cite{patterns}, we see that both
of these direct operators can only arise from loop corrections containing {\em
solely} non-standard virtual excitations, so, as was assumed in AEW, their
coefficients must be of order
\begin{equation}
\alpha_{\mu W} \sim g/(16 \pi^2) \quad {\rm and}
\quad \alpha_{\mu B} \sim g'/(16 \pi^2).
\label{estimates}
\end{equation}
As a result, these operators can give a contribution as large as \footnote{See
Eq.~(2.13) of AEW.}
\begin{equation}
a_\mu^{\rm direct} \sim 10^{-6}/\Lambda^2\qquad {\rm (\Lambda~in~TeV)}
\label{direct}
\end{equation}
$\Lambda$ is to be compared with the scale of weak symmetry breaking $v\approx
250~\gev.$  This suggests that, to account for the value reported\cite{BNL},
the scale of new physics could be rather large. However, this result is
deceptive.  In the SM, all fermion masses arise from Yukawa couplings to the
Higgs field. Setting the muon mass to zero is natural, in the technical sense
commonly employed, only if this gives rise to an additional chiral symmetry.
If one does not regard this additional global symmetry as accidental, then it
must also be a property of the underlying theory in this limit. Consequently,
we will assume that the muon mass is {\em naturally} light and that the
underlying theory, like the SM, has a chiral symmetry protecting the muon mass
in the limit that it is taken to be zero. Since these two direct operators
break that chiral symmetry, naturalness for the muon mass is intimately
connected with contributions to $\gmu2$, as was noted in AEW.  Naturality
requires that their coefficients contain a factor proportional to $y_\mu\sim
m_\mu/v\approx 4 \times 10^{-4}.$ This implies that the magnitude of the
coefficients of these operators in Eq.~(\ref{directo}) be reduced by such a
factor so that their contribution to the anomalous moment is of order 
\begin{equation}
a_\mu^{\rm direct} \sim 10^{-10}/ \Lambda^2 \qquad {\rm (\Lambda~in~TeV)}
\label{direct2}
\end{equation}
This suggests that the scale of new physics $\Lambda$ can not be much larger
than the weak scale $v$ itself, and therefore the associated particles may
well have masses of order of the SM vector bosons (100~\gev). This agrees with
the conclusions stated in~\cite{BNL} concerning the mass scale in certain
supersymmetric models, which are prototypical of models of this type.

Among the 79 remaining dimension-six operators in the SM, there are only a few
that could produce effects as large as observed.  Those must correspond to
vertices that both (a) can arise from some theory in tree approximation and
(b) can contribute to $\gmu2$ in one-loop order.  These are (in the notation
of Ref.~\cite{patterns}) 
\begin{eqnarray}
\hspace{-20pt} \ocal_{\phi l}^{(1)} &\equiv& {i (\phi^\dagger D_\mu \phi)
(\bar l
\gamma^\nu l)},\qquad
\ocal_{\phi l}^{(3)} \equiv {i (\phi^\dagger \tau^I D_\mu \phi)(\bar l
\gamma^\nu \tau^I l)},\hspace{-14pt}\cr
\ocal_{\phi e} &\equiv& {i (\phi^\dagger D_\mu \phi) (\bar e 
\gamma^\nu e)}.
\label{svf}
\end{eqnarray}
\vspace{-.3in}
\begin{eqnarray}
\ocal_{ll}^{(1)}&\equiv&{(\bar l \gamma_\nu l) (\bar l \gamma^\nu
l)},\quad\hfil 
\ocal_{ll}^{(3)}\equiv{(\bar l \gamma_\nu \tau^I l) (\bar l
\gamma^\nu\tau^I l)},\hspace{-3pt}
\cr
\ocal_{e e}&\equiv& { (\bar e \gamma_\nu e) (\bar e \gamma^\nu
e)},\quad\hfil
\ocal_{le} \equiv {(\bar l \gamma_\nu l)(\bar e \gamma^\nu e)}.\hfill
\label{Zprime}
\end{eqnarray}
\vspace{-.3in}
\begin{eqnarray}
\ocal_{lq}^{(1)}&\equiv&{(\bar l_\mu \gamma_\nu l_\mu) (\bar q
\gamma^\nu q)},
\quad\hfil
\ocal_{lq}^{(3)}\equiv{(\bar l_\mu \gamma_\nu \tau^I l_\mu) 
(\bar q \gamma^\nu \tau^I q)},\hspace{-3pt} \cr
\ocal_{\mu u}&\equiv&{(\bar \mu_R \gamma_\nu \mu_R) (\bar u \gamma^\nu
u)},
\quad\hfil
\ocal_{\mu d}\equiv{(\bar \mu_R \gamma_\nu \mu_R) (\bar d \gamma^\nu
d)},\cr
\ocal_{qd\mu}&\equiv&{(\bar l_\mu \gamma_\nu q)(\bar d \gamma^\nu
\mu_R)}.
\label{vectorlq}
\end{eqnarray}
\vspace{-.3in}
\begin{eqnarray}
\hspace{-10pt} \ocal_{lu}&\equiv&{(\bar l_\mu u)(\bar u l_\mu)},~ 
\ocal_{ld}\equiv{(\bar l_\mu d)(\bar d l_\mu)},~ 
\ocal_{q\mu}\equiv{(\bar q \mu_R)(\bar \mu_R q)}, \hspace{-23pt} \cr
\ocal_{lq} &\equiv& {(\bar l_\mu \mu_R) \epsilon (\bar q u)},~ 
\ocal_{lq}'\equiv {(\bar l_\mu u) \epsilon (\bar q \mu_R)}.
\label{scalarlq}
\end{eqnarray}
Here, $l (q)$ denotes a left-handed lepton (quark) doublet; $e (u,d)$, a
generic right-handed lepton (quark) singlet; and $\phi$, the SM Higgs doublet.
All these fermion fields could carry generational indices, and generation
mixing is in general permissible.  The $\tau^I$ are the Pauli matrices, and
$\epsilon\equiv i\tau^2$. We will consider each group in turn.  The operators
in Eq.~(\ref{svf}) modify the couplings of the $W^\pm$ and $Z^0$ to leptons of
order $\alpha_{\phi l}\sim g v^2/\Lambda^2$.  They affect $\gmu2$ very much
like the SM weak corrections, \ie, less than $10^{-9}$ times $\alpha_{\phi l}$
or $\alpha_{\phi \mu}$.  Quite aside from theoretical arguments, these non-SM
couplings are already constrained by LEP measurements\cite{gw} to be less than
about 1\%. Therefore, they change $\gmu2$ by less than $10^{-11}$ and can be
safely ignored.~\footnote{The presence of right-handed neutrinos would allow
another operator to the above list, {\it viz.,} $ i (\phi^T \epsilon D_\mu
\phi) (\bar \nu_R \gamma^\mu \mu_R) $ which would give rise to a right-handed
$W$ coupling. The experimental data on the $W$ branching ratios\cite{pdg},
however, insures that the corresponding contribution to $\gmu2$ is $ \lesim 5
\times 10^{-12} $ and can also be ignored.}

The four-fermion interactions (\ref{Zprime})--(\ref{scalarlq}) are also
constrained in various ways but cannot be excluded entirely. Each of the four-
lepton operators in Eq.~(\ref{Zprime}) could  arise from the exchange of a
heavy vector boson and would contribute to $\gmu2$ when two of the fields
correspond to muons.  Note that the first two operators involve only left-
handed couplings; the third, only  right-handed couplings; the fourth, both
left- and right-handed couplings.

At first sight, one might anticipate that operators involving muon fields of a
single chirality could not contribute to $\gmu2$, since the effective
operators Eq.~(\ref{directo}) involve a chiral flip of the muon.  However,
this is not correct, since a chirality-conserving, dimension-six operator such
as $(\bar l_{\mu}\sigma^{\alpha\beta} D\slash\;\l_\mu) B_{\alpha\beta}$ is, by
use of the classical equations of motion, equivalent to the corresponding
direct operator $\ocal_{\mu B}$ times the Yukawa coupling $y_\mu$ of the Higgs
field to the muon.\footnote{One attractive feature of such a chirally-coupled
$Z'$ is that it would always give a contribution to $\gmu2$ that is natural,
even if it were to cause transitions to a  heavy fermion.}

Returning to the discussion of the limits on the contributions of these
operators, there are already strong bounds on a generic $U(1)$ gauge boson, as
typically arises in models of unification, with $M_{Z'}\gesim 550~\gev$ if it
has coupling strengths comparable to SM couplings\cite{pdg}. Such vector
bosons would make a contribution to $\gmu2$ at least an order of magnitude
smaller than the SM weak correction. However, one can imagine a more exotic
$Z'$ that could be much lighter and account for the discrepancy observed.
Consider, for example, a $Z'$ that coupled only to leptons.\footnote{Such
models must concern themselves with anomaly cancelation, but we will assume
that can be accomplished.}  In general, generation-changing operators must
have small coefficients because they would lead to flavor-changing neutral
currents. However, a $Z'$ that did not cause transitions between generations
or that coupled essentially only to second-generation leptons cannot be
excluded. For example, a $Z'$ of mass 200~GeV that coupled to muons with three
times the strength of the SM $Z^0$ would account for the discrepancy observed.
Although such a $Z'$ would cause $Z^0\!-\!Z'$ mixing, the effects on the known
properties of the $Z^0$ and other observations can easily be seen to be
acceptably small.

If one performs a Fierz transformation on $\ocal_{le}$, one obtains the
equivalent operator $(\bar l e)(\bar e\,l)$.  Such an interaction could arise
from the exchange of a heavy scalar doublet.  The constraints on such a
particle are similar to those on the hypothetical $Z'$ discussed above.
Therefore, the most likely candidate to contribute to $\gmu2$ would be a
scalar that coupled diagonally or exclusively to second-generation leptons,
giving $(\overline{l_\mu}\mu_R)(\overline{\mu_R}l_\mu)$. Such a scalar doublet
with mass M and Yukawa coupling $h'$ would make a contribution to $\gmu2$ of
order $(m_\mu/v)(h'/4\pi M)^2$, which could be comparable to our previous
estimates of the magnitude of direct operators.  

Turning now to the two classes of operators involving both leptons and quarks,
Eqs.~(\ref{vectorlq}) and (\ref{scalarlq}), most could arise either from
various kinds of leptoquarks or from a $Z'$ that coupled to both leptons and
quarks.  However, unlike previously, in these modes such a $Z'$ cannot
contribute to $\g2$ at one-loop order, since its coupling to the photon arises
through mixing and therefore contributes only to the electric coupling of the
muon rather than to its magnetic coupling.  

The operator $\ocal_{lq}$ obviously could arise from the exchange of a
colorless scalar doublet that coupled to both quarks and leptons. However, as
with the $Z'$ just discussed, this scalar would be in the wrong channel to
contribute to $\g2$ in one-loop order.

Performing a Fierz transformation on the first four operators in
Eq.~(\ref{vectorlq}), one may rewrite them as 
\begin{eqnarray}
\ocal_{lq}^{(1)}&\equiv&{(\bar l_\mu \gamma_\nu q) (\bar q \gamma^\nu
l_\mu)},
\quad
\ocal_{lq}^{(3)}\equiv{(\bar l_\mu \gamma_\nu \tau^I q) (\bar q \gamma^\nu
\tau^I l_\mu)},
\hspace{-3pt}\cr
\ocal_{\mu u}&\equiv&{(\bar \mu_R \gamma_\nu u) (\bar u \gamma^\nu \mu_R)},
\quad
\ocal_{\mu d}\equiv{(\bar \mu_R \gamma_\nu d) (\bar d \gamma^\nu \mu_R)}.
%\cr
%\ocal_{qd\mu}&\equiv&{(\bar l_\mu \gamma_\nu \mu_R)(\bar d \gamma^\nu q)}.
\label{vectorlqtwo}
\end{eqnarray}
Thus, we see alternatively that all five operators of this set could  arise
from exchange of a vector leptoquark that is a color triplet having the weak
isospin and hypercharge assignments appropriate to each operator.  Similarly,
the operators in Eq.~(\ref{scalarlq}) except $\ocal_{lq}$ could arise from
scalar leptoquark exchanges with various electroweak quantum numbers.

Experimentally, current bounds on masses and couplings of leptoquarks imply
contributions of $O(10^{-11}) $ or less for leptoquarks that generate lepton-
flavor violation\cite{hera}. In other cases, the lower bounds on leptoquarks
are only around $200~\gev$\cite{pdg}, not so strong as to exclude them as
potential contributors.  Theoretically, in a weak coupling scenario,
leptoquarks sufficiently light to affect $\g2$ noticeably would be quite
unexpected. Typically, scalar and vector leptoquarks arise in models in which
quarks and leptons are unified, such as the Pati-Salam $SU(4)$ model or in
$SU(5)$ or $SO(10)$ GUTs.  The scales of unification in such models can be
estimated and are much too large to be relevant, typically $\gesim
10^{12}~\gev$ or more.\footnote{In contrast, if leptoquarks were composite,
they could have a significant effect on $\gmu2$.  See next section.}

Finally, the operators $\ocal_{lq}$ and $\ocal_{lq}'$ are obviously closely
related; indeed, if one performs a Fierz transformation on
$\ocal_{lq}$, has 
\begin{equation}
8\ocal_{lq}= -4 \ocal_{lq}' - 
\left( {\bar l_\mu} \sigma_{\alpha\beta} u\right)
\epsilon \left( {\bar q} \sigma^{\alpha\beta} \mu_R\right).
\end{equation}
Thus, in addition to arising from a scalar leptoquark, another possibility is
the exchange of a massive, antisymmetric tensor field, which is an alternative
way to describe a vector leptoquark.  These tensor excitations, if present,
should transform as weak isodoublets and couple to the fermions through a
$\sigma_{\mu\nu}$-type interaction. These constraints are problematic: the
fermion coupling violates the gauge symmetry required for the unitarity of the
free tensor Lagrangian\cite{townsend} and would naturally drive the tensor
mass to a value $ \gg \Lambda$. In addition the non-Abelian coupling appears
to be inconsistent with the condition of weak coupling\cite{tf}. For these
reasons, we will not consider this possibility further.

In sum, we conclude that, if the observed discrepancy is to emerge from an
underlying theory in which perturbation theory may be used, then its origin
must be from the direct operators Eq.~(\ref{directo}) or from models having a
rather exotic heavy vector boson or a leptoquark with the appropriate
properties.  We now turn to the consideration of the non-decoupling case, in
which the underlying model may involve strong interactions.

%%%%%%%%%%%%%%%%%%%%%%%%%%%%%%%%%%%%%%%%%%%%%%%%%%%%%%%
\bigskip
\paragraph{Non-Decoupling Case}

\def\qwe{f}

If the physics underlying the standard model does not decouple the power-
counting arguments are modified by the presumed strong-interaction effects. We
will consider the simplest realization of this possibility where the low-
energy theory contains only the Standard Model fields {\em without} a scalar
doublet. In this (chiral) theory symmetry breaking is achieved by introducing
a unitary field $U$, which replaces the SM doublet (this corresponds to the
case where the physical Higgs particle becomes heavy with $v$ kept fixed). In
the unitary gauge $U=1$, and the chiral Lagrangian reproduces the unitary-
gauge Standard Model Lagrangian without the terms containing the Higgs
excitation. For the present discussion  the relevant effective Lagrangian is
\begin{eqnarray}
&& \lcal \equiv { g \qwe_W \over \Lambda}\bar l_\mu \sigma^{\mu\nu}
\tau^I U r_\mu W^I_{\mu\nu}
 + { g' \qwe_B \over \Lambda} \bar l_\mu \sigma^{\mu\nu} B_{\mu\nu} U
r_\mu ; \hspace{-15pt}
\label{n.dec.l}
\end{eqnarray}
(plus Hermitian conjugate) where $r_\mu = (0,\mu_R)$. The naturality bound for
the coefficients (using naive dimensional analysis\cite{georgi}) is $ |
\qwe_{W,B} | \lesim 1 $. In addition these operators can modify the $\mu$
mass, a simple estimate shows that the requirement that this radiative
correction be small implies $| \qwe_{W,B} | e^2 \Lambda /(4\pi)^2 \ll m_\mu$
where $e$ denotes the proton charge. For this scenario, however, there is an
upper bound on the scale $\Lambda$ equal to $4 \pi v \sim 3~\tev$. This then
implies the  condition
\begin{equation}
|\qwe_{W,B}| \lesim 0.06
\end{equation}
This constraint can be naturally implemented if it assumed that the
coefficients take the form $\qwe_{W,B} \equiv (m_\mu/\Lambda)
\overline{\qwe_{W,B}}$.

The direct contributions to the anomalous magnetic moment are
\begin{eqnarray}
\delta a &=& 4 m_\mu (\qwe_W - \qwe_B )/ \Lambda \cr
&\lesim& 8.4 \times 10^{-6} \left| \qwe_{W,B}/ 0.06 
\right|(3 \hbox{TeV})/\Lambda
\end{eqnarray}
If the coefficients are of the form $\qwe_{W,B} \sim ( m_\mu/\Lambda) \times
O(1)$ we find $|\delta a | \sim ( 2 m_\mu/\Lambda)^2 $ $ \sim 4.6 \times 10^{-
9} $ for $ \Lambda \sim 3~\tev$. This estimate is significantly larger than
the standard model contribution but it is (perhaps coincidentally) comparable
to the deviation reported in the recent BNL experiment.

As with the decoupling scenario, there are non-direct contributions to $\gmu2$
generated by the operators Eq.~(\ref{svf}), and the same limits apply. In this
case, however, the naive-dimensional analysis estimate of the operator
coefficients is very different. In particular the 4-fermion operators
$\ocal_{lq},~\ocal_{lq}'$ are expected to appear with coefficients $ \sim
y_\mu/v^2 $ (assuming,as before, a natural realization of explicit chiral
symmetry breaking). This would also lead to a contribution as large as $10^{-
9}$ to $\gmu2$.

\paragraph{Conclusions}
To summarize, assuming that the scale of physics beyond the SM exceeds the
weak scale, we have considered all possible origins of deviations from the SM.
If the underlying theory can be treated perturbatively, then the only
possibilities to account for so large an effect would be models producing the
direct operators Eq.~(\ref{directo}), of which SUSY models are prototypical,
or models having unexpectedly light leptoquarks or having a rather exotic
heavy vector boson that couples predominantly to muons. The scale of new
physics must be sufficiently low that evidence of  most of these mechanisms
would be likely to appear in the new data anticipated from Collider Run II at
the Fermilab TeVatron. Other possible origins, such as a generic $Z'$ or
anomalous couplings of the SM vector bosons, either are already excluded by
existing experimental bounds or would have to be unnaturally large to account
for the magnitude of the observed discrepancy.  On the other hand, if the
underlying new physics involves a new strong interaction, as in technicolor
models, then the physics could be
accounted for by a variety of effective operators.  It is harder to pinpoint
the precise way to check this alternative, but evidence of compositeness on
this scale would also be likely to show up at the TeVatron.

\paragraph{Acknowledgments}
One of us (MBE) would like to thank G.~L. Kane for discussions concerning SUSY
models.  This work was supported in part by the U.S. Department of Energy.

\paragraph{Addendum}  
While this manuscript was being completed, a number of papers appeared
considering one or another of the alternatives touched on here.  These include
supersymmetric models\cite{susy}, an additional vector boson\cite{zprime},
leptoquarks\cite{lepto} as well as nondecoupling scenarios\cite{nondecouple}.
Consideration of models involving extended scalar sectors  has
begun\cite{multihiggs}. For a general theoretical overview, see
Ref.~\cite{review}.

%}  % end /large
%\vfill\eject


\begin{thebibliography}{99}

\bibitem{BNL}
H.~N.~Brown {\it et al.}  [Muon $\g2$ Collaboration],
hep-ex/0102017, submitted to Phys.\ Rev.\ Lett.\ .

%\cite{LEP}
\bibitem{LEP}
R.~Barate {\it et al.}  [ALEPH Collaboration],
Phys.\ Lett.\ {\bf B495}, 1 (2000) [hep-ex/0011045];
M.~Acciarri {\it et al.}  [L3 Collaboration],
Phys.\ Lett.\ {\bf B495}, 18 (2000) [hep-ex/0011043].

\bibitem{aew}
C.~Arzt {\it et al.}
Phys.\ Rev.\ {\bf D 49}, 1370 (1994) [hep-ph/9304206].

%\cite{LEPprecise}
\bibitem{LEPprecise}
See, for example, G.~Montagna, {\it et al.}
Riv.\ Nuovo Cim.\ {\bf 21N9}, 1 (1998) [hep-ph/9802302], 
and references therein.

%\cite{Buchmuller:1986jz}
\bibitem{Buchmuller:1986jz}
W.~Buchmuller and D.~Wyler,
Nucl.\ Phys.\ {\bf B268}, 621 (1986).

%\cite{patterns}
\bibitem{patterns}
C.~Arzt {\it et al.}
Nucl.\ Phys.\ {\bf B433}, 41 (1995) [hep-ph/9405214].

%\cite{weinberg}
\bibitem{weinberg}
S.~Weinberg,
``The Quantum Theory of Fields. vol. 2: Modern Applications,''
{\it  Cambridge, UK: Univ. Pr.,1996.}

\bibitem{gw}
B.~Grzadkowski and J.~Wudka,
Phys.\ Lett.\ B {\bf 364}, 49 (1995) [hep-ph/9502415].

\bibitem{pdg}
D.~E.~Groom {\it et al.}  [Particle Data Group Collaboration],
%``Review of particle physics,''
Eur.\ Phys.\ J.\ C {\bf 15}, 1 (2000);
references to the original literature are cited therein.

\bibitem{hera}
C.~Adloff {\it et al.}  [H1 Collaboration],
Eur.\ Phys.\ J.\ C {\bf 11}, 447 (1999) [hep-ex/9907002].

\bibitem{townsend}
P.~K.~Townsend,
%``Covariant Quantization Of Antisymmetric Tensor Gauge Fields,''
Phys.\ Lett.\ B {\bf 88}, 97 (1979).
S.~P.~de Alwis, {\it et al.}
%``Unitarity In Antisymmetric Tensor Gauge Theories,''
Phys.\ Lett.\ B {\bf 190}, 122 (1987).

\bibitem{tf}
D.~Z.~Freedman and P.~K.~Townsend,
Nucl.\ Phys.\ B {\bf 177}, 282 (1981).
T.~E.~Clark, {\it et al.}
Nucl.\ Phys.\ B {\bf 308}, 379 (1988).
S.~P.~de Alwis, {\it et al.}
Nucl.\ Phys.\ B {\bf 303}, 57 (1988).

%\cite{georgi}
\bibitem{georgi}
H.~Georgi,
Phys.\ Lett.\ {\bf B298}, 187 (1993) [hep-ph/9207278] and references
therein;
B.~Holdom,
Phys.\ Rev.\ {\bf D 56}, 7461 (1997) [hep-ph/9706527].

\bibitem{susy}
L.~Everett, {\it et al.}, hep-ph/0102145;
S.~Komine, {\it et al.}, hep-ph/0102204;
J.~L.\ Feng and K.~T.~Matchev, hep-ph/0102146;
J.~Ellis, {\it et al.}, hep-ph/0102331.

\bibitem{zprime}
D.~Choudhury {\it et al.}, hep-ph/0102199.

\bibitem{lepto}
K.~Cheung, hep-ph/0102238;
U.~Mahanta, hep-ph/0102176;
D.~Chakraverty {\it et al.}, hep-ph/0102180.

\bibitem{nondecouple}
K.~Lane, hep-ph/0102131;
Z.~Xiong and J.~M.~Yang, hep-ph/0102259;
P.~Das,{\it et al.}, hep-ph/0102242.

\bibitem{multihiggs}
A.~Dedes and H.~E.~Haber, hep-ph/0102297.
E.~Ma and M.~Raidal, hep-ph/0102255.

\bibitem{review}
A.~Czarnecki and W.~J.~Marciano,
hep-ph/0102122.

\end{thebibliography}
\end{document}